\documentstyle[aps,feynmp,epsf,twocolumn]{revtex}

\def\simlt{\stackrel{<}{{}_\sim}}
\def\simgt{\stackrel{>}{{}_\sim}}
\newcommand{\md}{m_\Delta}
\newcommand{\be}{\begin{eqnarray}}
\newcommand{\ee}{\end{eqnarray}}
\newcommand{\ben}{\begin{eqnarray*}}
\newcommand{\een}{\end{eqnarray*}}

\newcommand{\ov}{\overline}

\newcommand{\Gt}{\tilde{G}}
\newcommand{\sig}{\sigma_{\pi N}}
\newcommand{\J}[1]{J^{\pi N}_{#1}}
\newcommand{\Jp}[1]{J^{\pi\pi}_{#1}}
\newcommand{\Jn}[1]{J^{NN}_{#1}}
\newcommand{\Gp}[1]{\Gamma^{\pi N}_{#1}}
\newcommand{\Gn}[1]{\Gamma^{N\pi}_{#1}}
\newcommand{\GN}[1]{\Gamma^{NN}_{#1}}
\newcommand{\GP}[1]{\Gamma^{\pi\pi}_{#1}}
\newcommand{\Gpn}[1]{{\cal G}^{\pi N}_{#1}}
\newcommand{\Gnp}[1]{{\cal G}^{N\pi}_{#1}}

\newcommand{\Om}[1]{\Omega_{#1}}

\begin{document}

\draft
\title{\bf Dilepton and Photon Emission Rates from a Hadronic Gas II}

\author{{\bf James V. Steele}$^1$, {\bf Hidenaga Yamagishi}$^2$  and
{\bf Ismail Zahed}$^1$} 

\address{$^1$Department of Physics, SUNY, Stony Brook, New York 11794, USA;\\
$^2$4 Chome 11-16-502, Shimomeguro, Meguro, Tokyo, Japan. 153.}
\date{\today}
\maketitle

\begin{abstract}
We extend our recent analysis of the dilepton and photon emission rates
to the case of finite temperature and baryon density, within the context
of a density expansion. To leading order, the effects of the baryon density
are assessed using data (photon emission) or constraints from broken
chiral symmetry (dilepton emission). Next to leading order effects are
worked out, and their contribution qualitatively assessed. The opening
of the $\pi N$ cut causes the photon rate to saturate the empirical
photon yield of WA80, but may not be enough to fully account for the excess
low mass dileptons seen at CERES.
\end{abstract}
\pacs{}
\narrowtext

{\bf 1. Introduction}
\vskip .25cm

Recent relativistic heavy-ion collisions at CERN have reported an
excess of dileptons over a broad range of lepton invariant mass, below
the rho mass \cite{CERES,HELIOS}. A possible excess was also reported
in the direct photon spectrum \cite{WA80}. A rate departure from p-A
collisions may indicate some medium modifications in the hadronic
phase \cite{ALL}, an idea that has spurred considerable attention
lately \cite{BROWN,SHURYAK,HATSUDA,RAPP,WEISE}.

In a recent paper \cite{US1} (here after referred to as {\bf I}), we have
analyzed the dilepton and photon emission rates from a baryon free
hadronic gas using on-shell chiral reduction formulas \cite{MASTER}
in the context
of a density expansion. At pion density $n_{\pi}$ and temperatures of
the order of the pion mass $m_{\pi}$, the expansion parameter was
identified with $\kappa_\pi = n_\pi/2m_{\pi}f_{\pi}^2 \simlt 0.3$
\cite{US1}, where $f_{\pi}=93$ MeV is the pion decay constant. Some
enhancement was observed in the low mass dilepton and photon rates due
to the ``tails" in the vector and axial-vector correlators. This
enhancement is however insufficient to account for the empirical
excess when inserted in a conventional hydro-evolution of the fire
ball, and could be traced back to the detector cuts in transverse
momentum \cite{PRAKASH}. To account for the data, an order of
magnitude enhancement in the present bare rate is needed in the low
mass region from $2m_{\pi}$ to $4m_{\pi}$.

At present CERN energies, both the S-Au and Pb-Au experiments show
evidence of a sizable baryonic density in the mid-rapidity region
(about two to three times nuclear matter density $\rho_0$). Could it
be that such an enhancement is caused by the strong pion-nucleon
interactions?  Some recent investigations suggest so in the context of
some effective models of pion-nucleon dynamics and at specific
kinematics (back-to-back emission) \cite{RAPP,WEISE}.

In this letter we would like to investigate the role of a finite
baryon density in a hadronic gas using the approach presented in {\bf
I}, with an emphasis on constraints brought about by broken chiral
symmetry and experiments beyond the threshold region. Specifically, we
will use the comprehensive framework introduced in \cite{MASTER} in
the form of on-shell chiral reduction formulas. The outcome is a set
of general on-shell Ward-identities that could either be saturated by
experiment, or approximated near threshold by an on-shell
loop-expansion \cite{MASTER,USBIG}, and above threshold by
resonance saturation \cite{MASTER1}. The results conform with
the strictures of broken chiral symmetry, relativistic crossing and
unitarity. In this sense they are more than conventional dispersion
analysis. Both the photon and dilepton rates are subjected to the same
analysis due to their common dynamical origin, {\it albeit} with
different kinematics.

In the presence of baryons, we are still able to organize our
calculation into a density expansion as was done in {\bf I} for the
pions alone.  This allows us to separate out the important effects and
ascertain the corrections from the higher order terms.  In section 2,
we outline the character of this density expansion at finite
temperature and density. In section 3, we discuss the photon rates,
using data and chiral constraints. The effects of nucleons are
assessed using a relativistic one-loop chiral expansion without the
$\Delta$ and a chiral expansion with the $\Delta$ is evaluated in
section 4.  In section 5, we extend our discussion to the dilepton
emission rates to determine the enhancement due to nucleons.  We then
assess the magnitude of the next to leading order corrections in the
density expansion in section 6.  Our conclusions are summarized in
section 7.

\vskip .5cm
{\bf 2. Dilepton and Photon Rates}
\vskip .25cm

In a hadronic gas in thermal equilibrium, the rate ${\bf 
R}$ of dileptons produced in an unit four volume follows from the thermal 
expectation value of the electromagnetic current-current correlation 
function \cite{LARRY}. For massless leptons with momenta $p_1, p_2$,
the rate per unit invariant momentum $q =p_1+p_2$ is given by 
\be
\frac {d{\bf R}}{d^4q} = -\frac{\alpha^2}{6\pi^3 q^2}\,\,
\,\,{\bf W} (q)
\label{1}
\ee
where $\alpha =e^2/4\pi$ is the fine structure constant,
\be
{\bf W} (q) =\! \int\! d^4x \, e^{-iq\cdot x} \,
{\rm Tr}\! \left(e^{-({\bf H}-\mu \,{\bf N}-\Omega)/T} \,\,{\bf J}^{\mu}
(x){\bf J}_{\mu} 
(0)\right), 
\label{2}
\ee
$e{\bf J}_{\mu}$ is the hadronic part of the electromagnetic current,
${\bf H}$ is the hadronic Hamiltonian, $\mu$ the baryon chemical
potential, $\bf N$ the baryon number operator, $\Omega$ the Gibbs
energy, $T$ the temperature, and the trace is over a complete set of
hadron states. For leptons with mass $m_l$, the right-hand side of
(\ref{1}) is multiplied by
\ben
(1+\frac {2m_l^2}{q^2})(1-\frac{4m_l^2}{q^2})^{\frac 12}.
\een
Similarly, the rate for photons follows from (\ref{2}) at
$q^2=0$. Specifically, 
\be
q^0 \frac {d{\bf R}}{d^3 q} = -\frac{\alpha}{4\pi^2} 
\,\,{\bf W} (q)
\label{3}
\ee
Both (\ref{1}) and (\ref{3}) follow from the same current-current correlator
(\ref{2}), albeit for off-shell and on-shell photons respectively. For
consistency, both emission rates will be assessed simultaneously as was 
stressed in {\bf I}. 

{}From the spectral representation and symmetry, the rates may also be
expressed in terms of the absorptive part of the time-ordered function 
\be
{\bf W} (q) = \frac 2{1+e^{q^0/T}} \,\,{\rm Im}{\bf
W}^F(q)   
\label{4}
\ee
\ben
{\bf W}^F (q) = i\! \int\!\! d^4x \, e^{iq\cdot x} \,
{\rm Tr} \left(e^{-({\bf H}-\mu\, {\bf N} -\Omega)/T} \,\,T^*{\bf
J}^{\mu} (x){\bf J}_{\mu} (0)\right). 
\een
We observe that (\ref{4}) vanishes at $T=0$ even for non-zero baryon
chemical potential. A cold nuclear state can neither emit real nor
virtual photons, for otherwise it would be unstable.

For temperatures $T\simlt m_{\pi}$ and baryonic densities $n_N\simlt
3\rho_0$ we may expand the trace in (\ref{4}) using pion and nucleon
states\footnote{We neglect the anti-nucleon contribution since it is
highly suppressed for the temperatures and densities we
consider.}. Expanding the trace in terms of free pions and nucleons
with Feynman boundary conditions, and summing over disconnected pieces
lead to the density expansion 
\be
&&-i{\bf W}^F (q) = \langle 0| \,{\cal O}_{\bf J}(q) |0\rangle_{\rm conn.}
+\sum_a  \int\! d\pi \, \langle \pi | \, {\cal O}_{\bf J}(q) |\pi
\rangle_{\rm conn.}
\nonumber\\
&&{}+\sum_{s,I} \int\! dN \, \langle N | \, {\cal O}_{\bf J}(q) |N
\rangle_{\rm conn.} 
+ ...
\label{5}
\ee
with the nucleon fields carrying implicit spin and isospin
$N=N^I(p,s)$, the pion fields implicit isospin $\pi = \pi^a(k)$, and
\ben
{\cal O}_{\bf J}(q) = \int\! d^4x \, e^{iq\cdot x} \,T^* {\bf J}^{\mu}(x)
{\bf J}_{\mu} (0) .
\een
The phase space factors are 
\be
dN = &&\frac {d^3p}{(2\pi)^3} \frac 1{2E_p}
\frac 1{e^{(E_p-\mu)/T} +1}
\nonumber\\
d\pi = && \frac {d^3k}{(2\pi)^3} \frac {1}{2\omega_k} 
\frac 1{e^{\omega_k/T} -1}\nonumber
\ee
with the nucleon energy $E_p=\sqrt{m_N^2 +p^2}$ and the pion energy 
$\omega_k=\sqrt{m_{\pi}^2 +k^2}$.  The matrix elements in (\ref{5}) 
correspond to the forward scattering amplitudes of a real (photon rate) or
virtual (dilepton rate) photon with on-shell nucleons and pions. 
We consider the nucleons to be in thermal equilibrium and ignore the
off-equilibrium effects caused by the collision dynamics 
on the mid-rapidity nucleons.  The chemical potential $\mu$
is fixed by specifying the nucleon density.  We observe that the first
term in (\ref{5}) when used in conjunction with (\ref{4}) is
reminiscent of the resonance gas approximation \cite{weldon}.
The density expansion is an expansion in the number of
pions and nucleons in the final state, which means by detailed balance
and time-reversal all possible thermal reactions in the initial
state.  

To leading order in the nucleon and pion densities $n_N=4\int dN\;
2E_p\;$ and $n_\pi=3\int d\pi\;2\omega_k\;$ only the first few terms
in (\ref{5}) contribute. A slight refinement is to note that the
reduction of one pion is associated with a factor of $1/f_{\pi}$, and
that of a nucleon is associated with a factor of $g_A/f_{\pi}$, where
$g_A =1.26$ is the nucleon axial charge. The dimensionless expansion
parameters should be $\kappa_\pi =n_\pi/2m_{\pi} f_{\pi}^2$ and
$\kappa_N= n_N\, g_A^2/2m_N f_{\pi}^2$. For $T\simlt m_{\pi}$ and
$n_N\simlt 3\rho_0$, we have $\kappa_\pi\sim \kappa_N\simlt 0.3$. The
truncation should be reasonable, unless severe infrared divergences
develop. This is unlikely in the hadronic phase since most resonances
are massive with finite widths.

\vskip .5cm
{\bf 3. Adding Nucleons}
\vskip .25cm

A full treatment of the first two terms of eq.~(\ref{5}) was carried out
in {\bf I}.  There the electromagnetic current was decomposed into an
isovector part ${\bf V}^3$ and an isoscalar part ${\bf B}$.  The {\bf
BB} and {\bf BV} correlators are expected to be small in the pionic
states, but in the presence of nucleons they must be reassessed.  The
vacuum contribution is proportional to ${\bf\Pi}_V$, the transverse part of
the vector correlator $\langle 0| T^* {\bf VV}|0\rangle$ which follows
from electroproduction data.  Also including terms to first order 
in the density expansion, we find
\be
&&{\rm Im} \,{\bf W}^F (q) = -3q^2 {\rm Im} \;{\bf \Pi}_V(q^2)
\nonumber\\
&&{}+\frac1{f_\pi^2}\int\! d\pi\, {\bf W}^F_\pi(q,k) 
+ \int\! dN\, {\bf W}^F_N (q,p) 
\nonumber\\
&&{}+ {\cal O}\!\left(\kappa_\pi^2, \kappa_N^2, \kappa_\pi\kappa_N
\right). 
\label{6}
\ee
The term linear in pion density can be reduced by the use of chiral
reduction formulas to a form amenable to experimental determinations.
That analysis showed that, to within a few percent, the only
important contributions were \cite{US1}
\be
&&{\bf W}^F_{\pi}(q,k) \simeq 12q^2 {\rm Im}\; {\bf \Pi}_V(q^2)
\nonumber\\
&&{}-6(k+q)^2 {\rm Im}\; {\bf \Pi}_A\left( (k+q)^2 \right) +
(q\rightarrow -q)
\nonumber\\
&&{}+8((k\cdot q)^2-m_\pi^2 q^2) {\rm Im}\;
{\bf \Pi}_V(q^2) 
\nonumber\\
&&\qquad\times{\rm Re} \left( \Delta_R(k+q) + \Delta_R(k-q) \right)  
\label{7}
\ee
with $\Delta_R (k)$ the retarded pion propagator 
\ben
\Delta_R (k) = {\bf PP} \frac 1{k^2-m_{\pi}^2} -i\pi {\rm sgn } (k^0 )
\delta (k^2 -m_{\pi}^2 ),
\een
and ${\bf \Pi}_A$ the transverse part of the axial correlator $\langle
0 | T^* {\bf j}_A {\bf j}_A| 0\rangle$ which follows from tau decay
data.  

\begin{figure}
\begin{center}
\leavevmode
\epsfxsize=3.375in
\epsffile{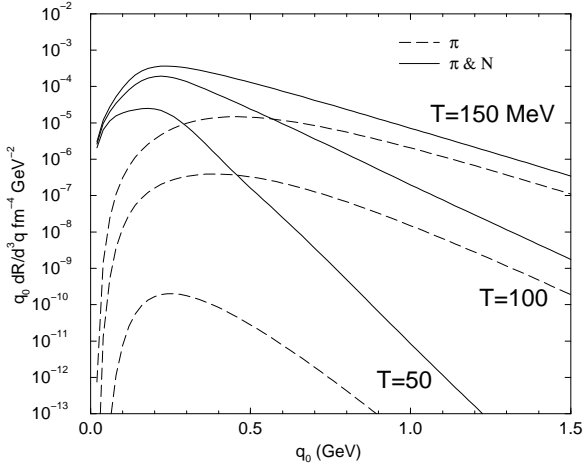}
\end{center}
\caption{\label{nucl.ps}
The photon emission rate to first order in the density
expansion for pions only (dashed) and both pions and nucleons
(solid).  A fixed nucleon density of $\rho_0$ was used.}
\end{figure}

Now taking into account the nucleons gives the third term, which is
just the spin-averaged forward Compton scattering amplitude on the
nucleon with virtual photons.  This is only measured for various
values of $q^2 \le 0$.  However, the dilepton and photon rates require
$q^2 \ge 0$.  Therefore, only the photon rate for this term can be
determined directly from data by use of the optical theorem
\be
e^2 {\bf W}^F_N(q,p) =
-4(s-m_N^2) \sum_I \sigma^{\gamma N}_{\rm tot.}(s)
\label{8}
\ee 
with $s=(p+q)^2$.  Experimentally, the isospin sum of the cross
section in (\ref{8}) can just be replaced by $\sigma^{\gamma d}_{\rm
tot.}$ \cite{landolt} due to the weak collective effect of the two
nucleons in the deuteron.  The total cross section of $\gamma d$ is
dominated by the $\Delta(1232)$ \cite{pdg}.  The result for the
instantaneous photon rate is shown in fig.~\ref{nucl.ps} for
$T=50,100,150$ MeV and a nucleon density of $n_N=\rho_0$.  From there
it is easy to see that the nucleon contribution dominates over the
pions by about an order of magnitude.  If the same situation were
true for the dilepton rate, the enhancement would probably be enough
to explain the excess dileptons in the low-mass region.  

This huge correction to the pion result near threshold is due to the
opening of the $\Delta$ channel.  Approximating the cross section in
(\ref{8}) by a narrow $\Delta$ resonance, we see the dimensionless
parameter of importance is $(\md-m_N)/2\Gamma_\Delta\sim 1$ which is
multiplied by $\kappa_N$ under thermal integration.  This enhancement
from the opening of a new threshold does not iterate in the low mass
region.  This ensures the density expansion is under control.  In
section 5 we check whether this enhancement in the photon rate is
enough to surpass the upper bound set by WA80 \cite{WA80}.

For off-shell photons, we must resort to chiral constraints to
determine the nucleon contribution to the dilepton rate.  Broken chiral
symmetry dictates uniquely the form of the strong interaction
Lagrangian (at tree level) for spin $\frac12$. Perturbative unitarity follows
from an on-shell loop-expansion in $1/f_{\pi}$, that enforces 
current conservation and crossing symmetry \cite{USBIG}. 

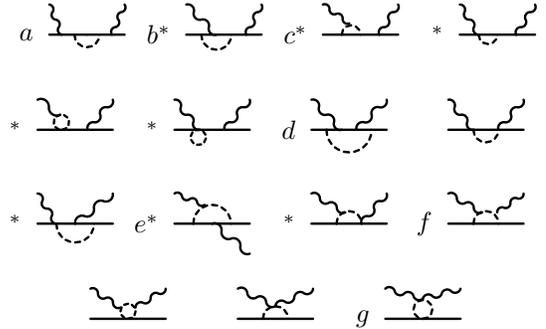
\begin{figure}[h]
\begin{center}
\leavevmode
\begin{fmffile}{mpcompton}
\fmfset{dash_len}{1.5mm}
\hbox{
\begin{fmfgraph*}(10,10)
 \fmfleft{i}
 \fmfright{o}
 \fmftop{c1,c2}
 \fmf{plain}{i,v1,v2,v22,v3,v4,o}
 \fmf{photon,tension=0}{c1,v1}
 \fmf{dashes,right,tension=0}{v2,v3}
 \fmf{photon,tension=0}{c2,v4}
 \fmflabel{$a$}{i}
\end{fmfgraph*}
\qquad
\begin{fmfgraph*}(10,10)
 \fmfleft{i}
 \fmfright{o}
 \fmftop{c1,c2}
 \fmf{plain}{i,v1,v2,v3,v4,o}
 \fmf{photon,tension=0}{c1,v2}
 \fmf{dashes,right,tension=0}{v1,v3}
 \fmf{photon,tension=0}{c2,v4}
 \fmflabel{$b^*$}{i}
\end{fmfgraph*}
\qquad
\begin{fmfgraph*}(10,10)
 \fmfleft{i}
 \fmfright{o}
 \fmftop{c1,c2}
 \fmf{plain}{i,v1,v2,v3,o}
 \fmf{photon,tension=0}{c1,v}
 \fmfforce{(.35w,.65h)}{v}
 \fmf{dashes,left,tension=0}{v1,v2}
 \fmf{photon,tension=0}{c2,v3}
 \fmflabel{$c^*$}{i}
\end{fmfgraph*}
\qquad
\begin{fmfgraph*}(10,10)
 \fmfleft{i}
 \fmfright{o}
 \fmftop{c1,c2}
 \fmf{plain}{i,v1,v2,v3,o}
 \fmf{photon,tension=0}{c1,v1}
 \fmf{dashes,right,tension=0}{v1,v2}
 \fmf{photon,tension=0}{c2,v3}
 \fmflabel{${}^*$}{i}
\end{fmfgraph*}
}

\medskip

\hbox{\qquad\quad
\begin{fmfgraph*}(10,10)
 \fmfleft{i}
 \fmfright{o}
 \fmftop{c1,c2}
 \fmf{plain}{i,v1,v2,o}
 \fmf{photon,tension=0}{c1,v}
 \fmf{photon,tension=0}{c2,v2}
 \fmfforce{(0.3w,0.7h)}{v}
 \fmf{dashes,right,tension=0}{v1,v}
 \fmf{dashes,right,tension=0}{v,v1}
 \fmflabel{${}^*$}{i}
\end{fmfgraph*}
\qquad
\begin{fmfgraph*}(10,10)
 \fmfleft{i}
 \fmfright{o}
 \fmftop{c1,c2}
 \fmf{plain}{i,v1,v2,o}
 \fmf{photon,tension=0}{c1,v1}
 \fmf{photon,tension=0}{c2,v2}
 \fmf{dashes,right,tension=0}{v1,v}
 \fmf{dashes,right,tension=0}{v,v1}
 \fmfforce{(0.3w,0.3h)}{v}
 \fmflabel{${}^*$}{i}
\end{fmfgraph*}
\qquad
\begin{fmfgraph*}(10,10)
 \fmfleft{i}
 \fmfright{o}
 \fmftop{c1,c2}
 \fmf{plain}{i,v1,v2,v3,v4,o}
 \fmf{photon,tension=0}{c1,v2}
 \fmf{photon,tension=0}{c2,v3}
 \fmf{dashes,right,tension=0}{v1,v4}
 \fmflabel{$d$}{i}
\end{fmfgraph*}
\qquad
\begin{fmfgraph}(10,10)
 \fmfleft{i}
 \fmfright{o}
 \fmftop{c1,c2}
 \fmf{plain}{i,v1,v2,o}
 \fmf{photon,tension=0}{c1,v1}
 \fmf{photon,tension=0}{c2,v2}
 \fmf{dashes,right,tension=0}{v1,v2}
\end{fmfgraph}
}

\medskip

\hbox{\qquad\quad
\begin{fmfgraph*}(10,10)
 \fmfleft{i}
 \fmfright{o}
 \fmftop{c1,c2}
 \fmf{plain}{i,v1,v2,v3,o}
 \fmf{photon,tension=0}{c1,v1}
 \fmf{photon,tension=0}{c2,v2}
 \fmf{dashes,right,tension=0}{v1,v3}
 \fmflabel{${}^*$}{i}
\end{fmfgraph*}
\qquad
\begin{fmfgraph*}(10,10)
 \fmfleft{i}
 \fmfright{o}
 \fmftop{c1,d1}
 \fmfbottom{d2,c2}
 \fmf{plain}{i,v1,v2,v3,o}
 \fmf{photon,tension=0}{c1,v}
 \fmfforce{(0.4w,0.75h)}{v}
 \fmf{photon,tension=0}{c2,v2}
 \fmf{dashes,left,tension=0}{v1,v3}
 \fmflabel{$e^*$}{i}
\end{fmfgraph*}
\qquad
\begin{fmfgraph*}(10,10)
 \fmfleft{i}
 \fmfright{o}
 \fmftop{c1,c2}
 \fmf{plain}{i,v1,v2,o}
 \fmf{photon,tension=0}{c1,v}
 \fmfforce{(0.45w,0.65h)}{v}
 \fmf{dashes,left,tension=0}{v1,v2}
 \fmf{photon,tension=0}{c2,v2}
 \fmflabel{${}^*$}{i}
\end{fmfgraph*}
\qquad
\begin{fmfgraph*}(10,10)
 \fmfleft{i}
 \fmfright{o}
 \fmftop{c1,c2}
 \fmf{plain}{i,v1,v2,o}
 \fmf{photon,tension=0}{c1,v11}
 \fmf{photon,tension=0}{c2,v22}
 \fmfforce{(0.45w,0.65h)}{v11}
 \fmfforce{(0.65w,0.65h)}{v22}
 \fmf{dashes,left,tension=0}{v1,v2}
 \fmflabel{$f$}{i}
\end{fmfgraph*}
}

\medskip

\hbox{\qquad\qquad\quad
\begin{fmfgraph}(10,10)
 \fmfleft{i}
 \fmfright{o}
 \fmftop{c1,c2}
 \fmf{plain}{i,v,o}
 \fmf{photon,tension=0}{c1,v1}
 \fmf{photon,tension=0}{c2,v2}
 \fmfforce{(0.4w,0.65h)}{v1}
 \fmfforce{(0.6w,0.65h)}{v2}
 \fmf{dashes,right,tension=0}{v,vv}
 \fmf{dashes,right,tension=0}{vv,v}
 \fmfforce{(0.5w,0.7h)}{vv}
\end{fmfgraph}
\qquad
\begin{fmfgraph}(10,10)
 \fmfleft{i}
 \fmfright{o}
 \fmftop{c1,c2}
 \fmf{plain}{i,v1,v2,o}
 \fmf{photon,tension=0}{c1,v}
 \fmf{photon,tension=0}{c2,v}
 \fmfforce{(0.5w,0.67h)}{v}
 \fmf{dashes,left,tension=0}{v1,v2}
\end{fmfgraph}
\qquad
\begin{fmfgraph*}(10,10)
 \fmfleft{i}
 \fmfright{o}
 \fmftop{c1,c2}
 \fmf{plain}{i,v,o}
 \fmf{photon,tension=0}{c1,v1}
 \fmf{photon,tension=0}{c2,v1}
 \fmfforce{(0.5w,0.77h)}{v1}
 \fmf{dashes,right,tension=0}{v,vv}
 \fmf{dashes,right,tension=0}{vv,v}
 \fmfforce{(0.5w,0.75h)}{vv}
 \fmflabel{$g$}{i}
\end{fmfgraph*}
}
\end{fmffile}

\caption{\label{diagrams.ps} 
One-loop diagrams for Compton scattering.  The graphs with a star also
have a mirror image diagram which must be taken into account and all
graphs except for those in the last line require the addition of a
crossed diagram.  The lettered graphs are discussed in Appendix A.}
\end{center}
\end{figure}

In general, after summing over spin, only four invariant functions
which depend on $s$ and $q^2$ remain: 
\be
&&\sum_s i\int\! d^4x e^{iq\cdot x} \langle N_{\rm out}(p)| T^* {\bf
J}_\mu(x) {\bf J}_\nu(0) | N_{\rm in}(p) \rangle
\nonumber\\
&&\qquad = 4\left( g_{\mu\nu} A + q_\mu q_\nu B + q_{(\mu} p_{\nu)} C
+ p_\mu p_\nu D \right).
\label{9}
\ee 
The absorptive part, which is of relevance here, only starts to
contribute at one-loop.  The four invariant functions are related by
current conservation --- leaving only $A$ and $B$ as independent.
To one-loop in $1/f_{\pi}$ the diagrams that contribute are
shown in fig.~\ref{diagrams.ps} and require the addition of the 
crossed diagrams as well.  The analytic expressions are quoted 
in Appendix A for arbitrary $q^2$. For on-shell photons, they fulfill
the non-renormalization condition $A(m_N^2,0)=Z$ with $Z=0(1)$ for the
neutron (proton).  This is fully given by the tree level nucleon
contribution, leaving $A^{\rm loops}(m_N^2,0)=0$ which has been
checked to hold analytically for the loop functions.

\begin{figure}
\begin{center}
\leavevmode
\epsfxsize=3.375in
\epsfysize=3.375in
\epsffile{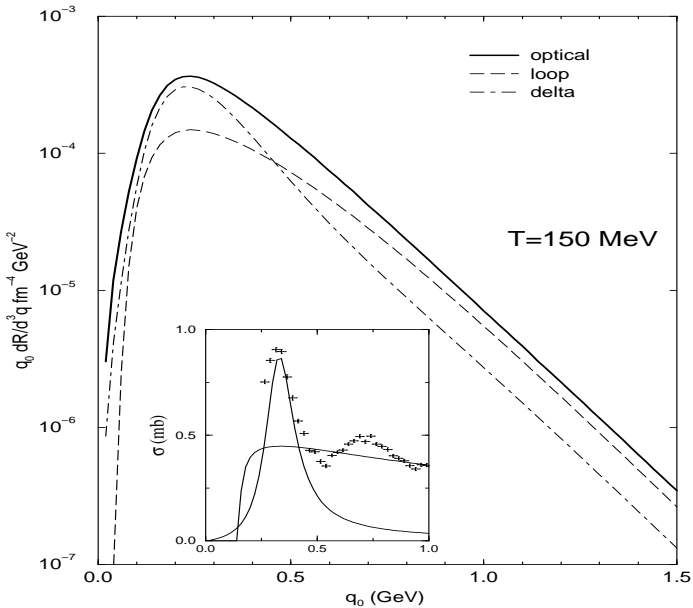}
\end{center}
\caption{\label{photon.ps}
The photon emission rate as in fig.~1 using data and the optical
theorem (solid), using the one-loop results (dashed), and using a
contribution from the $\Delta$ (dashed-dotted).  A fixed nucleon 
density of $\rho_0$ was used.  The inset shows data for
$\sigma^{\gamma d}_{\rm tot.}$ in millibarns as a function of lab
momentum in GeV from [17] and the good agreement for the
imaginary parts of the one-loop and delta results discussed in the
text.}
\end{figure}

The results for the photon rate are shown in fig.~\ref{photon.ps} for
a representative temperature $T=150$ MeV (results are similar for
other temperatures).  Comparison is made with the nucleon result from
fig.~\ref{nucl.ps}.  The one-loop amplitude reproduces the shape
attained from the data quite well --- even for larger photon energies
--- with the exception of the small bump around the $\Delta$ peak.
That missing feature, however, is kinematically around the region of
interest for the enhancement of the dilepton data ($300$ MeV above
threshold).  Therefore to make a full statement about the nucleon
contribution for finite $q^2$, we will include the $\Delta$ in the
analysis.

\vskip .5cm
{\bf 4. The Role of the $\Delta$}
\vskip .25cm

Chiral symmetry does not constrain the form of the strong interaction
Lagrangian at tree level for spin $\frac32$ in the relativistic
approximation. We therefore treat the $\Delta$ as a resonance in the
continuum, and use time reversal invariance and data to constrain the
form of the transition matrix element.  We decompose the $N$-$\Delta$
vector transition matrix element in general as\footnote{We could also
include terms with $\gamma_\mu u^\mu$, but these are suppressed near
the mass-shell, where we are interested.}
\be
\langle N(p) | {\bf V}^a_\mu | \Delta(k) \rangle &=& \overline{u}(p)
\Bigg[ Q(q^2) (\gamma_\mu q_\nu - g_{\mu\nu} \rlap/q)
\nonumber\\
&&{}+\left( R(q^2) + \rlap/q \overline{R}(q^2) \right) (q_\mu q_\nu -
g_{\mu\nu} q^2 )
\nonumber\\
&&{}+iS(q^2) \sigma_\mu^\lambda q_\lambda q_\nu \Bigg] \gamma_5
u^{\nu,a}(k) 
\label{delta}
\ee
with $q=k-p$.  Current conservation dictates the specific tensor form
of (\ref{delta}).  Note that the isoscalar part of the electromagnetic
current does not contribute to the nucleon-$\Delta$ transition matrix 
element.

The modulus square and sum over all spins and isospins of
(\ref{delta}) will be denoted by ${\cal M}(k^2,q^2)$ and is quoted in
full in Appendix B.  It is directly related to the $\Delta$
contribution of the forward scattering amplitude ${\bf W}^F_N$,
\ben
{\bf W}^F_N(q,p) = {\rm Im} \frac{4m_N\md}{s-\md^2+i\md\Gamma_\Delta}
{\cal M}(s,q^2) + (s \to u)
\een
with $u=(p-q)^2$.  If we were not to take the imaginary part, the full
tree level result must also include the nucleon Born terms.  At
$q^2=0$ this is equal to the isospin sum of $8A(s,0)$ and the same
non-renormalizability condition of the last section requires
$A^\Delta(m_N^2,0)=0$.  This is satisfied trivially by the
decomposition used in eq.~(\ref{delta}), since the charge was fixed to
zero by current conservation. 

If we do not contract the vector indices of ${\bf W}^F_N$ as in
eq.~(\ref{9}), we can also determine $B(s,q^2)$.  The
polarizabilities are defined as
\ben
&&\overline{\alpha} + \overline{\beta} =-2\alpha m_N A''(m_N^2,0) 
= \frac{8\alpha}9 \frac{m_N}{\md^2} \frac{\md^2+m_N^2}{\md^2-m_N^2} Q^2
\\
&&\overline{\beta} = -\frac{\alpha}{m_N} B(m_N^2,0) = \frac{8\alpha}9
\frac{Q^2}{\md -m_N} ,
\een
over-constraining the value of $Q$.  Although experiment has shown the
electric polarizability dominates over the magnetic one
$\overline{\alpha} > \overline{\beta}$, the $\Delta$ is a magnetic
dipole effect and tends to overestimate $\overline{\beta}$ in other
calculations \cite{bkm}.  Therefore we concentrate on fixing a
reasonable value for $\overline{\alpha} +\overline{\beta} =
10\times10^{-4}$ fm$^3$.  This indeed gives a $\overline{\beta} \simeq
14\times10^{-4}$ fm$^3$ about three times the experimental value.  This
corresponds to $Q(0)=2.75/m_N$.  Translating into our notation, it is
comparable to the value $Q\simeq 2.5/m_N$ obtained by a different
method \cite{deltapapers}.  

The decay width of the $\Delta$ is given by
\ben
\Gamma_{\Delta\to N\gamma} = \alpha m_N \frac{\md^2-m_N^2}{8\md^2} {\cal
M}(\md^2,0) \simeq 0.72 \mbox{MeV},
\een
and can be used to fix $S(0)=1.2/m_N^2$.  This compares well with
$S\simeq 1.0/m_N^2$ from \cite{deltapapers}.  

The form factors $R, \overline{R}$ do not contribute at $q^2=0$.  For
finite $q^2$, the $\Delta$ contribution is mostly around its mass shell and
so we can use the Dirac equation to rewrite $\rlap/q \overline{R} \to
(\md + m_N) \overline{R}$ in (\ref{delta}) and absorb $\overline{R}$
into the definition of $R$.  Then using the vertex (\ref{delta}) in
pion electroproduction,  $R$ can be shown to be proportional to the
longitudinal part of the cross section.  This is on the order of
$10\%$ of the total result even for moderate $q^2$ and therefore
we will take $R=\overline{R}=0$ for the rest of this paper.

\begin{figure}
\begin{center}
\leavevmode
\epsfxsize=3.375in
\epsffile{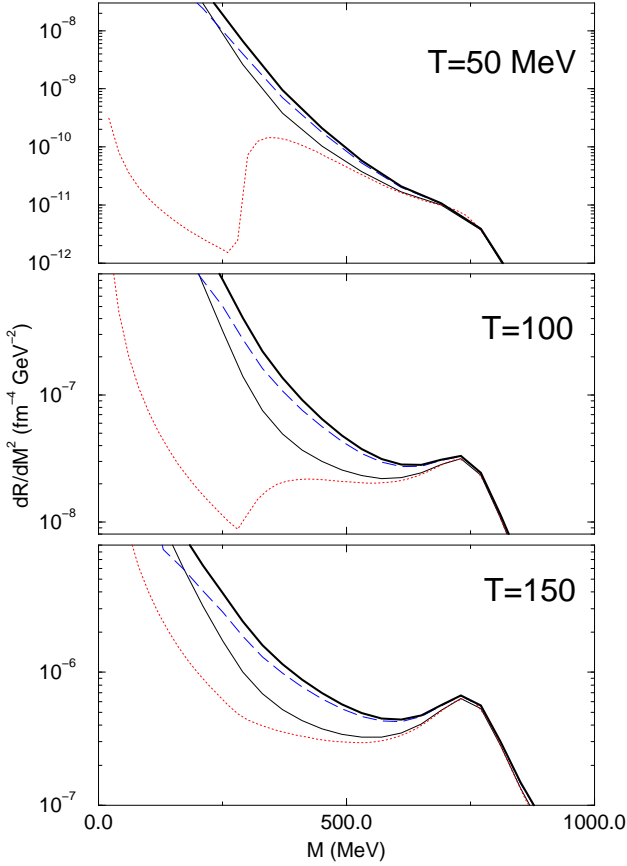}
\end{center}
\caption{\label{dilep.ps}
The dielectron rate for pions alone (dotted), pions and $\Delta$ (solid),
and pions and one-loop (dashed).  The contribution from pions,
$\Delta$, and one-loop together is represented by the thick solid
line.  A fixed nucleon density of $\rho_0$ was used.} 
\end{figure}

We plot the photon rate due to the $\Delta$ along with the one-loop
result in fig.~\ref{photon.ps}.  From there we see that the $\Delta$
rate does quite well around the peak and then dies away for larger
energies.  In fact, adding the one-loop and $\Delta$ rates, we come
very close to reproducing the nucleon result obtained from the data.
This even includes the secondary peak, seen in the inset of
fig.~\ref{photon.ps}, which is fit moderately well by the loop
contribution. Therefore we take both contributions into account as a
way to parameterize the photon spectral function with on-shell  chiral
symmetry, relativistic crossing, and unitarity constraining the
function near threshold, and experiment constraining it above.  We now
can move to finite $q^2$ to ascertain the dilepton rate.

\vskip .5cm
{\bf 5. Dilepton Rates with Nucleons}
\vskip .25cm

We will neglect the momentum dependence of the form factors in
eq.~(\ref{delta}) since they are expected to be of monopole form
$(1-q^2/4M_+^2)^{-1}$ with $M_+=\frac12(m_N+\md)$ \cite{dipole} and
hardly change for the $q^2$ we will examine (less than 1 GeV$^2$).
Then taking the results of the previous section to $q^2>0$ gives the
curves in fig.~\ref{dilep.ps}.  Also shown is the one-loop result and
the pion result from {\bf I} for comparison.

Qualitatively, the rates are similar to the photon case.  The enhanced
bump of the $\Delta$ is smeared out and enhances the rate by as much
as a factor of ten (for $T=100$ MeV) before gradually dying out again
near the $\rho$ peak. The tail of the one-loop goes above the $\Delta$
result around $M=200$~MeV. This may be easily understood by noting
that the absorptive part involves the $\pi N$ cut, with pions
typically carrying a momentum of order $m_{\pi}$ and the nucleon a
momentum of order $p_F\sim m_{\pi}$.  The one-loop is also overtaken
by the pion result at the $\rho$ peak. A closer examination of the
one-loop real part shows that it is about half of the tree result for
$q^2\simgt (300 \rm{MeV})^2$, above which the $1/f_\pi$ expansion
is no longer reliable. Fortunately, the energetic 
dilepton pairs with $q^0\simgt 300$ MeV are thermally suppressed by 
almost an order of magnitude at the highest temperature, typically
$e^{-300/150}\sim 1/7$.

\begin{figure}
\begin{center}
\leavevmode
\epsfxsize=3.375in
\epsffile{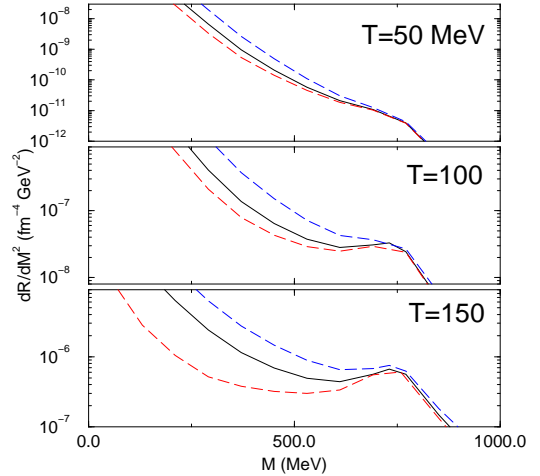}
\end{center}
\caption{\label{dens.ps}
The dielectron rate including the pions, $\Delta$, and one-loop for
$\rho_0$ (solid), $\frac12\rho_0$ (lower dashed), and $3\rho_0$ (upper
dashed).} 
\end{figure}

Combining the one-loop and $\Delta$ results gives the thick solid line
in fig.~\ref{dilep.ps}.  The dominant effect in our case comes from
the continuum and not the $\Delta$ resonance.  At $M=400$ MeV, the
inclusion of nucleons enhances the rate by a factor of three.  Others
who have taken nucleons into account through various methods
\cite{BROWN,RAPP,WEISE} find enhancements in the rate similar to our
result.  However, the details of the in-medium effects are not needed
in this analysis and so should be considered more general.  Also in
our calculations there is no shift of the dilepton pair production
$\rho$ peak.

In fig.~\ref{dens.ps} we show the dependence of the $T=150$ MeV
dielectron rate on the nucleon density for $n_N=\frac12 \rho_0,
\rho_0,$ and $3\rho_0$.  At $T=150$ MeV and $\frac12 \rho_0$, the
result is dominated by the pions as the nucleon contribution is
extremely small.  This shows the effect is truly from the nucleon
density.  Even so, only at $3\rho_0$ does the rate flatten out at the
value of the $\rho$ peak rather than increasing towards it.  

In order to fully understand the role of the cuts, we follow
\cite{RAPP} and evolve the instantaneous rate over space-time using
recent transport equation results \cite{BROWN} and experimental
cuts \cite{CERES}.  Assuming a homogeneous expansion, the volume and
temperature only depend on time and can be parameterized as
\[
V(t) = V_0 \left( 1+ \frac{t}{t_0} \right)^3 \qquad
T(t) = \left( T_i-T_\infty \right) e^{-t/\tau} + T_\infty
\]
with $t_0=10 (10.8)$ fm/c and $\tau=8 (10)$ fm/c for S-Au (Pb-Au)
collisions.  The freeze-out time is $t_{\rm f.o.}=10 (20)$ fm/c.  We
absorb $V_0$ into an overall normalization constant that includes the
charged particle distribution as well.  RQMD predicts the initial baryon
density $n_B\simeq 2.5\rho_0$ for S-Au and $4\rho_0$ for Pb-Au
\cite{SORGE} which translates into $n_N\simeq 0.7\rho_0 (1.0\rho_0)$.
Only the pion and nucleon densities are retained since we are
expanding the emission rates (\ref{5}) in terms of {\it stable}
final states, summing over all possible initial states. This point
is particularly transparent in the $\rho$ region. Assuming chemical
equilibrium gives a constant nucleon chemical potential.  The rate
then can be written as
\ben
\frac{(dN/d\eta\, dM)}{(dN_{ch}/d\eta)} 
= N_0 M \int_0^{t_{\rm
f.o.}} \! dt\; V(t) \int\! \frac{d^3q}{q_0} A(q^0,q^2)\;
\frac{d{\bf R}}{d^4q} 
\een
with the normalization constant $N_0=6.76 (1.33)\times 10^{-7}$ fixed
by the transport results in \cite{BROWN}.  The acceptance function $A$
ensures the detector cuts at CERES ($p_\perp > 200 (175)$ MeV, $2.1<
\eta < 2.65$, and $\Theta_{ee} > 35$ mrad) are taken into account.
The finite mass resolution is also taken into account by folding the
spectrum with a Gaussian averaging function given by the CERES
collaboration \cite{PRAKASH}.

\begin{figure}
\begin{center}
\leavevmode
\epsfxsize=3.175in
\epsffile{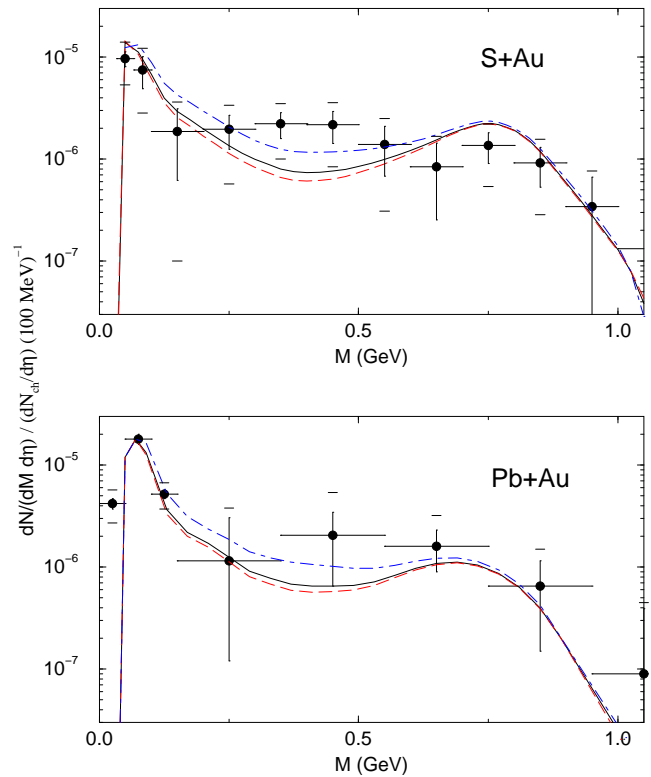}
\end{center}
\caption{\label{data.ps} Our dielectron rate including the $\Delta$
and one-loop contributions evolved in space-time as in [8] for S-Au
and Pb-Au collisions.  In the upper graph, $n_N=0,0.7\rho_0$, and
$2.5\rho_0$ are plotted as the dashed, solid, and dashed-dotted lines
respectively.  In the lower graph, the lines are for
$n_N=0,1.0\rho_0$, and $4\rho_0$.  The data are from [1].  The
systematic errors are added linearly to the statistical error bars
to give the cross line. The Pb-Au data is preliminary.}
\end{figure}

Finally, we plot the results for S-Au and Pb-Au collisions in
fig.~\ref{data.ps}.  This requires the addition of the Dalitz and
omega decays which were obtained from the transport model
\cite{BROWN}.  The data is presented by adding the statistical and
systematic errors linearly.  The Pb-Au data is still preliminary.  For
$n_N=0$, the pion result alone has already been analyzed in two
hydro-dynamical models \cite{PRAKASH,SHURYAK}.  Comparison of this alone
can be used to test the validity of the evolution in space-time
described above. Our result is shown as the dashed line in
fig.~\ref{data.ps}.  The overall shape due to the cut looks
reasonable.  The values for the rate in \cite{PRAKASH} are an order of
magnitude smaller, but this can be traced back to a shorter time in
the hadronic phase.  The agreement with \cite{SHURYAK} are fair.  We can
think of the results of this space-time evolution as an upper bound on
the results from a true hydro-dynamical model.

Adding the nucleon contribution gives the solid line in
fig.~\ref{data.ps}.  The effect of the cuts is dramatic, resulting in
a very small enhancement.  Only if we take the extreme case of the
baryon density totally saturated by nucleons do we start to reach the
lower error bars of the data in the $M=200-400$ MeV regime as shown by
the dashed-dotted line.  The large effect seen in fig.~\ref{dilep.ps}
is not present because the temperature dies away quickly, thereby
decreasing the nucleon density and rate dramatically as seen in
fig.~\ref{dens.ps}.   

Comparing to \cite{RAPP}, our rate is slightly smaller in the region
of interest.  This can be traced back to the bare rates.  For $M\simgt
400$ MeV, their rate for $n_N=0.5\rho_0$ is {\em larger} than the rate
for $n_N=\rho_0$ \cite{RAPP1} which differs from our result in which
the nucleon contribution dies away continuously leaving only the pion
contribution (fig.~\ref{dens.ps}).  This is related to the depletion
of the $\rho$ peak in \cite{RAPP}, which does not occur in our case.

\begin{figure}
\begin{center}
\leavevmode
\epsfxsize=3.175in
\epsffile{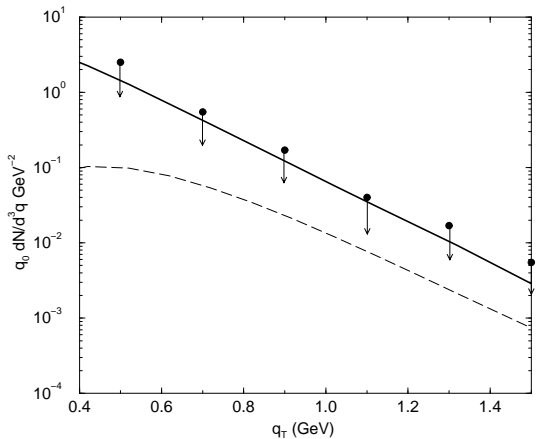}
\end{center}
\caption{\label{photdata.ps} Our photon rate with (solid) and without
(dashed) the nucleon contribution discussed in this paper.  The data
are upper bounds from [3].}
\end{figure}

We can also evolve the photon rates and compare with the upper bounds
set by WA80 for S-Au\cite{WA80}.  The procedure is similar to that for
the dileptons above.  The acceptance cut is $2.1\le \eta \le 2.9$ and
no integration over momentum is required.  In this case the factor
$N_0$ is only the volume $V_0$ which we normalize by comparison with
\cite{PRAKASH} for $n_N=0$.  This fixes $V_0\simeq 300$ fm$^3$ which
corresponds to an initial radius of $4$ fm and agrees with central
collision estimates.  The result for the pion (dashed line) and
nucleons for the extreme case of $n_N=2.5\rho_0$ (solid line) are
shown in fig.~\ref{photdata.ps}.  The inclusion of nucleons put the
rate right on the edge of the upper limit for the data.  Since we have
analyzed both the dilepton and photon rates simultaneously, this
implies that more enhancement of the dilepton rate would overshoot the
photon data, a particularly important point in our analysis.

\vskip .5cm
{\bf 6. Higher Order Terms}
\vskip .25cm

We have now assessed all the terms of eq.~(\ref{5}) linear in the
density.  The result is that they cannot explain the CERES data.  In
general, higher order terms will be suppressed by more powers of the
expansion parameter.  This implies the inclusion of higher order
effects will not be dramatic.  However, this rule of thumb can be
upset if new thresholds open up in the low mass region.  Typically,
the second order corrections to eq.~(\ref{6}) are of the form
\be
\delta\; {\rm Im}\, {\bf W}^F(q) &\simeq& \frac1{2f_\pi^4} \int\! d\pi_1 \,
d\pi_2\, {\bf W}^F_{\pi\pi}(q,k_1,k_2)
\nonumber\\
&&{}+\frac1{f_\pi^2} \int\! d\pi\, dN\, {\bf W}^F_{\pi N}(q,k,p)
\nonumber\\
&&{}+\frac12\int\! dN_1\, dN_2\, {\bf W}^F_{NN}(q,p_1,p_2) 
\label{higher}
\ee
with ${\bf W}^F_{\alpha\beta}$ referring to the forward scattering for
real or virual photons $\gamma^* \alpha\beta\to \gamma^*\alpha\beta$
as used above.  Note that matrix elements involving nucleons carry an
implicit factor of $1/f_\pi^2$ since the contribution only starts at
one-loop.  Experimentally, we can estimate the strength of the third
term in eq.~(\ref{higher}) at $q^2=0$ as follows.  We noted in
connection with eq.~(8) in the last section that the cross section for
Compton scattering on a deuteron is identical to the sum of Compton
scattering on a proton and neutron individually.  This seems to
indicate that the collective effect of two nucleons is small and we
will ignore it for the rest of this paper.

The $\Delta$ may be thought of as a bound state of a pion and a
nucleon, and so the second term in eq.~(\ref{higher}) could be
important.  We also must retain the isoscalar contribution ${\bf
B}_\mu$ to the electromagnetic current when nucleons are involved.  We can
still reduce out the pions as given in \cite{MASTER} if we assume the
pions do not interact with the isoscalar part.  This is true in the
soft pion limit and so should be a good approximation \cite{US1}.
The overall form of ${\bf W}^F_{\pi N}$ is very similar to ${\bf
W}^F_\pi$, in which one pion was also reduced out, with the vacuum
matrix elements being replaced by averages in a nucleon state.  The
result is
\ben
&&{\bf W}^F_{\pi N}(q,k,p) = -4\;{\rm Im}\, i \sum_{s,I} \int\! d^4x\;
e^{iq\cdot x}
\\
&&\times \langle N_{\rm out}(p) | T^* {\bf J}_\mu(x) {\bf J}^\mu(0) - {\bf
B}_\mu(x) {\bf B}^\mu(0) | N_{\rm in}(p) \rangle
\\
&&{}+\sum_{a,\{s,I\}}  \epsilon^{a3g} \epsilon^{a3f}
\\
&&\times {\rm Im} \bigg( g^{\mu\nu} - (k+q)^{(\mu} (2k+q)^{\nu)}
\Delta_R(k+q) 
\\
&&{}+ (k+q)^\mu (k+q)^\nu (2k+q)^2 \Delta^2_R(k+q) \bigg)
\\
&&\times\int\! d^4x\; e^{i(k+q)\cdot x} i \langle N_{\rm out}(p) | T^* {\bf
j}_{A\mu}^f(x) {\bf j}_{A\nu}^g(0) | N_{\rm in}(p) \rangle
\\
&&+(q\rightarrow -q)
\\
&&{}+8 k^\mu k^\nu {\rm Im}\, i\sum_{s,I}\int\! d^4x\; e^{iq\cdot x}
\\
&&\times{\rm Re}\left( \Delta_R(k+q) + \Delta_R(k-q) \right)  
\\
&&\times\langle N_{\rm out}(p) | T^* {\bf J}_\mu(x) {\bf J}_\nu(0) - {\bf
B}_\mu(x) {\bf B}_\nu(0) | N_{\rm in}(p) \rangle
\\
&&{}+3m_\pi^2 f_\pi \int\!\! d^4x d^4y \; e^{iq\cdot(x-y)}
\\
&&\times \sum_{s,I}
{\rm Im}\; \langle N_{\rm out}(p) | T^* {\bf J}_\mu(x) {\bf J}^\mu(y)
\hat{\sigma}(0) | N_{\rm in}(p) \rangle 
\\
&&{}-k^\alpha k^\beta \int\!\! d^4x\, d^4y\, d^4z \; e^{iq\cdot(x-y)}
e^{-ik\cdot z}
\\
&&\times {\rm Im}\; i\langle N_{\rm out}(p) | T^* {\bf J}_\mu(x) {\bf
J}^\mu(y)  {\bf j}^a_{A\alpha}(z) {\bf j}^a_{A\beta}(0) | N_{\rm
in}(p) \rangle 
\\
&&{}+k^\beta {\rm Im}\;
\left(\delta^\alpha_\mu - (2k+q)_\mu (k+q)^\alpha \Delta_R(k+q) \right) 
\\
&&\times \int\!\! d^4x\, d^4y \; e^{ik\cdot(y-x)} e^{-iq\cdot x}
\\
&&\times\sum_{a,\{s,I\}} i\epsilon^{a3e}
\langle N_{\rm out}(p) | T^* {\bf j}^e_{A\alpha}(x) {\bf j}^a_{A\beta}(y) 
{\bf J}^\mu(0) | N_{\rm in}(p) \rangle
\\
&&{}+(q \rightarrow -q) + (k \rightarrow -k) + (q, k \rightarrow -q, -k)
\een
with $\tilde{\bf \Pi}_A$ denoting the transverse part of $\langle N|
T^* {\bf j}_A {\bf j}_A | N\rangle$ summed over spin and isospin.
We use the same analysis as the last section for the ${\bf JJ}$ term.
The contribution here is negative, but weighted by an extra factor of
$\kappa$. It is about 10\% of the leading contribution.  Since there
is no prominent resonance with the quantum numbers of the isoscalar
term ${\bf BB}$, the imaginary part should also be small and will be
neglected.  

The second term contains the axial current ${\bf j}_A$, which is a
pionic contribution, and so this term may feed into the 
$\Delta$.  Using the decomposition ($k=p'-p$)
\ben
\langle N(p) | {\bf j}_{A\mu}^a | \Delta(p') \rangle &=&
\overline{u}(p) \Bigg[ F(k^2) g_{\mu\nu} + G(k^2) \gamma_\mu k_\nu
\\
&&{}+H(k^2) k_\mu k_\nu + i I(k^2) \sigma_\mu^\lambda k_\lambda k_\nu
\Bigg] u^{\nu,a}(p') 
\een
an analysis of $\pi N$ scattering in \cite{kacir} showed that
$F(0)=1.38$ and $G(0)=0.4/m_N$. Using the fact that $H$ and $I$ are
suppressed for large $N_c$, we can justify keeping only the $F$ term
to find 
\ben
&&\int\! d^4x\; e^{iK\cdot x} i\langle N_{\rm out}(p) | T^* {\bf
j}_{A\mu}^f(x) {\bf j}_{A\nu}^g(0) | N_{\rm in}(p) \rangle =
\\
&&= -\frac{32F^2}{9} \delta^{fg} \frac{p\cdot K + m_N (m_N +
\md)}{(p+K)^2 -\md^2+i\md\Gamma_\Delta} 
\\
&&\times\left( g_{\mu\nu} - \frac{(p+K)_\mu (p+K)_\nu}{\md^2} \right)
+ (K\to -K)
\een
where only the $\Delta$ channel has been retained. This term and its
crossed counterpart gives a positive contribution to the rate.
Numerical results show this term is appreciable right near threshold
but quickly shrinks to $10\%$ of the terms linear in $\kappa$ and
practically vanishes even before the $\Delta$ peak is reached.  The
damping is simply coming from the exponential factors in phase space
and so the only place where this term is significant is within the
Dalitz tail.  This behavior is expected to occur for all terms in
${\bf W}^N_{\pi N}$ and so we may neglect the terms of order
$\kappa_N\kappa_\pi$.

For completeness, we quote the result for the first term in
(\ref{higher}) in Appendix C.  Qualitatively, we note that the full
$\pi\pi$ scattering amplitude as well as terms of photon-pion
scattering similar to ${\bf W}_\pi$ appear with an additional
suppression factor of $\kappa$.  In the soft pion limit most of the
correlation functions in the pionic state are amenable to correlations
in the vacuum, some of which were assessed in {\bf I} and found to be
small. Hence, we expect qualitatively that ${\bf W}^F_{\pi\pi}$ is
suppressed by the factor $\kappa_\pi^2$ as expected and hence are
small in comparison with the lower order terms.

\newpage
{\bf 7. Conclusions}
\vskip .25cm

We have extended the analysis in {\bf I} to account for possible
nucleon densities in assessing the photon and emission rates from a
hot hadronic gas.  We have used the very general framework put
forward in \cite{MASTER,MASTER1,USBIG} to allow for a comprehensive analysis
of the emission rates in powers of the matter densities, and enforce
the known constraints of broken chiral symmetry on-shell, current
conservation, and relativistic unitarity. We believe that most
approaches to this problem as well others involving issues of chiral
symmetry in matter, have to embody the chiral Ward-identities
established on-shell in \cite{MASTER}, whether at threshold or above.

Our purpose of this paper was to use these arguments in combination
with a density expansion to account for a systematic analysis of the
dilepton and photon rates from a hot hadronic gas at finite nucleon
density. For the photon rate, we have used the constraints from broken
chiral symmetry and data to leading order in the densities. A large
enhancement of photons in the low energy region was found. Using an
on-shell expansion to one-loop and a general decomposition of the
$\Delta$, we were able to reproduce the photon rates at $q^2=0$. 

We have used the same amplitude to estimate the dilepton production
rate at $q^2>0$. In the latter, we have found an enhancement from the
$\pi N$ cut in the low mass region for a nucleon density on the order
of nuclear matter density on top of the already large enhancement from
thermal reactions {\it alone}. This enhancement is caused by the
opening of a threshold in the $\pi N$ channel, and does not iterate
coherently in the next order corrections, which we have found to be
qualitatively small and mostly confined to the Dalitz region.

For S-Au and Pb-Au reactions, we have evolved the dilepton rate in
space-time and taken the experimental cuts.  The result is not enough
to explain the data, even with some liberal choices of the parameters.
The nucleon densities do trigger large dilepton yields but die out
quickly after short reaction times.  In addition, we have evolved the
photon rates and found the results on the edge of the upper bounds of
the data in the extreme case.  This means that any further enhancement
would overshoot the empirical measurements. Since we have a controlled
expansion scheme, these results lead us to suspect the nucleons may not
explain the data.  Any mechanism which could explain the data would
need to have a very specific form in order not to upset the photon
yield.  However, this space-time evolution is still only a qualitative
assessment.  Quantitative results from a full hydro-dynamical model
which takes into account the quark-gluon phase, flow, and hadronic
multiplicities is needed for a final analysis.

Our results imply a different interpretation for the role of the
nucleons compared with a number of recent estimates
\cite{RAPP,WEISE,NORENBERG}.  Our analysis, however, is different in a
number of ways. We have evaluated the rates at finite temperature,
since there is no emission in cold matter. Also our results go
smoothly into those of {\bf I} at zero nucleon density, thereby
enforcing the known constraints of (on-shell) broken chiral symmetry,
current conservation, relativistic crossing and unitarity for all
dilepton kinematics.  It is important to stress that our density
expansion is an expansion in the number of pions and nucleons in the
final state, which means by detailed balance and time-reversal all
possible thermal reactions in the initial state.  Finally, our
qualitative assessment of the next to leading order shows that the
expansion is reliable in the low mass region, with no apparent need
for coherent resummations.

The empirical consequence of these differences is that the dilepton
strength in the $\rho$ region from a hot hadronic gas remains about
constant and dominated by thermal $\pi\pi$ pairs.  The enhancement
below this region in invariant mass does not come at the expense of
this peak.  A consequence of this is the enhancement is very sensitive
to the temperature and nucleon density and dies away quickly as the
fireball evolves.  The resolution in the $\rho$ region for the CERES
data is not enough to sort out the dilepton emission from `cold'
$\omega$'s. An improvement on this resolution maybe an important step
in assessing the effects of interaction as reported here in comparison
to those discussed in \cite{BROWN,RAPP,WEISE,NORENBERG}. Another
improvement, as stressed in {\bf I}, is to have more empirical
correlations between the photon and dilepton yields.

\vglue 0.6cm
{\bf \noindent  Acknowledgements \hfil}
\vglue 0.4cm
We would like to thank G.E.~Brown, G.Q.~Li, M.~Prakash, E.V.~Shuryak, and
H.~Sorge for discussions.  We especially thank R.~Rapp for
discussion and supplying us with the details of his calculation.  This
work was supported in part by the US DOE grant DE-FG02-88ER40388.

\vskip .5cm
{\bf Appendix A}
\vskip .25cm

The on-shell one-loop result can be written in terms of Feynman parameter
integrals.  Since we are only considering the imaginary parts, the
(real) subtraction constants will be ignored.  
\ben
&&f_\pi^2 A(s,q^2) = Z(g_A^2-1) \left( 1- \frac{q^2}{s-m_N^2} \right)
\Jp{22}(q^2) 
\\
&&{}-4m_N ( g_A \Gt + \sig) \ov{\GP{3}}(q^2)
\\
&&{}+ Z \frac{q^2G^2}{s-m_N^2} \Bigg[ 2\Gp{3}(s) - \Jn{}(q^2) 
-4\Gn{3}(s) 
\\
&&\qquad{}+ \frac32 \J{1}(s) - m_\pi^2 \Gp{}(s) \Bigg]
\\
&&{}-3Z^2 G^2 \frac{4q^2m_N^2}{(s-m_N^2)^2} \left[ \J{}(s) -
\J{1}(s) \right] 
\\
&&{}+4G^2 \left[ \GP{3}(q^2) - \GN{3}(q^2) \right] + G^2 \Jn{}(q^2)
\\
&&{}+\frac12 Z G^2 \left[ 4\GN{3}(q^2) - \Jn{}(q^2) - 3\J{1}(s)\right]
\\
&&{}+G^2\left( 1 - \frac32 Z \right) (s-m_N^2) \Gp{}(s)
\\
&&{}+ 2ZG^2 (s-m_N^2) \Gp{1}(s) + 2ZG^2 q^2 \Gp{2}(s) 
\\
&&{}+(Z-2)2m_\pi^2 G^2 \Gpn{4} + (Z-1)4m_\pi^2 G^2 \Om{4}
\\
&&{}- 4m_\pi^2 G^2 \Gnp{4}
+\frac12(Z-2)G^2 q^2 \left[ m_\pi^2 \Gpn{} + 2\GN{12}(q^2)\right] 
\\
&&{}+{\rm crossed}
\een
\ben
&&f_\pi^2 B(s,q^2) = -4\left(m_N\sig-(g_Am_N-G)^2\right)
\\
&&\qquad\qquad\times\Bigg[ \GP{456}(q^2)
-\GP{12}(q^2) + \frac14 \GP{}(q^2) \Bigg] 
\\
&&{}+Z G^2 \frac{4m_N^2}{s-m_N^2} \Bigg[ 2 \Gp{4}(s) + 2\Gp{5}(s) -
\Gn{}(s)
\\
&&\qquad{}+ 4\Gn{1}(s) + 2\Gn{2}(s) - 4\Gn{4}(s) - 4\Gn{5}(s) \Bigg]
\\
&&{}+2(Z-2)G^2 \left[ \GN{456}(q^2) - \GN{12}(q^2) \right]
\\
&&{}+2ZG^2 \Gp{456}(s) -2G^2 \Gp{12}(s) - G^2 \left[ \Gn{}(s)
-2\Gn{12}(s)\right] 
\\
&&{}-2ZG^2 \left[ 2\Gn{456}(s) - \Gn{12}(s) \right]
\\
&&{}+2(Z-2)m_\pi^2 G^2 \left[ \Gpn{10}-\Gpn{3} \right] + 4(Z-1)
m_\pi^2 G^2 \Om{5}
\\
&&{}-4m_\pi^2 G^2 \left[ \Gnp{10}- \Gnp{3} + \frac14 \Gnp{} \right]
\\
&&{}+{\rm crossed}
\een
\ben
C = \frac{2}{q^2 + m_N^2-s} \left( A + q^2 B \right) \qquad
D = \frac{2q^2}{q^2 + m_N^2-s}\; C
\een
with $\Gamma_{456} = \Gamma_4 + 2\Gamma_5 + \Gamma_6$ and $\Gamma_{12}
= \Gamma_1 + \Gamma_2$.  Also $g_A$ is the nucleon axial charge,
$\sig$ is the pion-nucleon sigma term, $G=f_\pi g_{\pi NN}$, and
$\Gt=2G-g_Am_N$.  The values used were $g_A=1.26$, $m_N=940$ MeV,
$m_\pi=140$ MeV, and $\sig=45$ MeV to ensure the nucleon is on-shell.
Taking $q^2=0$ and $\sig\to0$ reduces the loop result to that given in
\cite{bkm}.  Conformity with the Ward-identities \cite{MASTER}
requires an on-shell expansion \cite{USBIG} as taken into account
here.

Each of the $J, \Gamma, {\cal G}$ and $\Omega$'s are Feynman
parameterization integrals quoted in full in \cite{USBIG}.  The
imaginary parts have cuts for $s>s_0=(m_N+m_\pi)^2$ and $q^2>
4m_\pi^2$.  The cuts for $q^2>4m_N^2$ do not impact this paper and are
not quoted, thereby making the contributions from $\Jn{}$ and $\GN{i}$
zero. The labeled diagrams in fig.~\ref{diagrams.ps}a-h are dominated
by $\J{}$, $\Gp{}$, $\Gn{}$, $\Gpn{}$, $\Om{}$, $\Gnp{}$, $\GP{}$, and
$\Jp{}$ respectively (fully given by these if the $\pi N$ coupling
were a constant rather than proportional to a derivative in the
Lagrangian). 

We only need to quote a subset of the integrals and
obtain the others by use of algebraic manipulations as outlined in
\cite{bkm}.  Using $\lambda(x,y,z)=x^2+y^2+z^2-2xy-2xz-2yz$,
$\lambda_\pi=\lambda(s,m_N^2,m_\pi^2)$,
$\lambda_q=\lambda(s,m_N^2,q^2)$, and $\lambda_{\pi q} =
\lambda(q^2,m_\pi^2,m_\pi^2)$, the imaginary parts for $q^2 \ge 0$
are 
\ben
&&{\rm Im}\; \Jp{}(q^2) = \frac{\sqrt{\lambda_{\pi q}}}{16\pi q^2} \;
\theta(q^2-4m_\pi^2) 
\\
&&{\rm Im}\; \J{}(s) = \frac{\sqrt{\lambda_\pi}}{16\pi s}\; \theta (s-s_0)
\\
&&{\rm Im}\; \GP{}(q^2) = \frac1{16\pi \lambda_{\pi q}}\;\theta(q^2-4m_\pi^2) 
\\
&&{\rm Im}\; \Gp{}(s) = - \frac1{16\pi\sqrt{\lambda_q}} \ln \left[
\frac{a_{\pi q} + \sqrt{\lambda_\pi \lambda_q}}{a_{\pi q} -
\sqrt{\lambda_\pi \lambda_q}} \right] \theta(s-s_0)
\\
&&{\rm Im}\; \Gn{}(s) = - \frac1{16\pi\sqrt{\lambda_q}} \ln \left[
\frac{b_{\pi q} + \sqrt{\lambda_\pi \lambda_q}}{b_{\pi q} -
\sqrt{\lambda_\pi \lambda_q}} \right] \theta(s-s_0)
\\
&&\qquad{}+\frac1{16\pi\sqrt{\lambda_q}} \ln \left[ \frac{c_{\pi q} +
\sqrt{\lambda_{\pi q} \lambda_q}}{c_{\pi q} - \sqrt{\lambda_{\pi q}
\lambda_q}} \right] \theta(q^2-4m_\pi^2)
\\
&&{\rm Im}\; \Gpn{} = \frac{\sqrt{\lambda_\pi}}{4\pi} \frac{s}{a_{\pi
q}^2 - \lambda_\pi \lambda_q} \;\theta(s-s_0)
\\
&&{\rm Im}\; \Gnp{} = \frac{\sqrt{\lambda_\pi}}{4\pi} \frac{s}{b_{\pi
q}^2 - \lambda_\pi \lambda_q} \;\theta(s-s_0)
\\
&&\qquad{}- \frac1{8\pi q^4} \frac{\sqrt{\lambda_{\pi
q}}}{s-m_N^2+m_\pi^2}\; \theta(q^2-4m_\pi^2)
\\
&&{\rm Im}\; \Om{} = \frac1{16\pi\sqrt{\lambda_q}(s-m_N^2-q^2)}\;
\theta(s-s_0)
\\
&&\times\left\{ \ln \left[ \frac{e_{\pi q} + \sqrt{\lambda_\pi
\lambda_q}}{e_{\pi q} - \sqrt{\lambda_\pi \lambda_q}} \right]
+ \ln \left[ \frac{f_{\pi q} + \sqrt{\lambda_\pi
\lambda_q}}{f_{\pi q} - \sqrt{\lambda_\pi \lambda_q}} \right] \right\}
\\
&&+ (s\to u)
\\
&&{}+ \frac1{16\pi\sqrt{\lambda_q}(s-m_N^2-q^2)}\;
\theta(q^2-4m_\pi^2)
\\
&&\times\left\{ \ln \left[ \frac{d_{\pi q} + \sqrt{\lambda_{\pi q}
\lambda_q}}{d_{\pi q} - \sqrt{\lambda_{\pi q} \lambda_q}}
\right]
+ \ln \left[ \frac{m_\pi^2q^2 + \sqrt{\lambda_q\lambda_{\pi
q}}}{m_\pi^2q^2 - \sqrt{\lambda_q \lambda_{\pi q}}} \right] \right\}
\een
with 
\ben
&&a_{\pi q} = (s-m_N^2+q^2)(s+m_N^2-m_\pi^2) -2sq^2
\\
&&b_{\pi q} = (s+m_N^2-q^2)(s+m_N^2-m_\pi^2) -2s(2m_N^2-m_\pi^2)
\\
&&c_{\pi q} = q^2(s-m_N^2-q^2+2m_\pi^2)
\\
&&d_{\pi q} = q^2(s-m_N^2-q^2-m_\pi^2)
\\
&&e_{\pi q} = (s-m_N^2+q^2)(s-m_N^2+m_\pi^2) -2sq^2
\\
&&f_{\pi q} = 2s(s-m_N^2) - (s-m_N^2+q^2)(s-m_N^2+m_\pi^2) .
\een
These imaginary parts reduce to those in \cite{bkm} for $q^2=0$.
Increasing $q^2$ leads to $\lambda_q$ becoming negative and the
analytically continued forms of the above expressions, for example
\ben
{\rm Im}\; \Gp{}(s) = - \frac1{8\pi\sqrt{-\lambda_q}} \arctan\left(
\frac{\sqrt{\lambda_\pi \lambda_q}}{a_{\pi q}}\right) \theta(s-s_0),
\een
must be used.

\vskip .5cm
{\bf Appendix B}
\vskip .25cm

We use the following conventions for the $\Delta$'s sum over spins and
propagator 
\ben
\sum_s u^i_\mu(p) \overline{u}^j_\nu(p) = \left( \delta^{ij} -
\frac{\tau^i\tau^j}3 \right) \frac{\rlap/p + \md}{2\md} P_{\mu\nu}(p)
\\
\Delta^{ij}_{\mu\nu}(p) = \left( \delta^{ij} - \frac{\tau^i\tau^j}3
\right) \frac{i(\rlap/p + \md)}{p^2-\md^2+i\md \Gamma_\Delta}
P_{\mu\nu}(p)
\een
with
\[
P_{\mu\nu}(p) = g_{\mu\nu} - \frac13 \gamma_\mu \gamma_\nu -
\frac1{3\md} \gamma_{[\mu} p_{\nu]} - \frac2{3\md^2} p_\mu p_\nu.
\]
Defining $(\nu,\nu_p,\nu_q) = (p\cdot q, k\cdot p,k \cdot q)/m_N$ and
$k^2=s$, we find
\ben
&&\sum_{s,I} | \langle N(p) | {\bf V}^3_\mu | \Delta(k) \rangle |^2
\equiv {\cal M}(s,q^2)
\\
&&\equiv \frac{8}{9\md} \Bigg[ 2m_N\nu_q^2 {\cal M}_1(s,q^2) +
2m_N\nu_q^2 q^2 {\cal M}_2(s,q^2)
\\
&&\qquad{}+q^2 {\cal M}_3(s,q^2) + q^4 {\cal M}_4(s,q^2) + q^6 {\cal
M}_5(s,q^2) \Bigg]
\een
\ben
&&{\cal M}_1(s,q^2) = \left( \frac{\nu}{\nu_q} + \frac{\nu_p r^2}{m_N}
\right) Q^2 + 2r^2 \nu_q\nu S^2 
\\
&&\qquad{}-2r(2\nu-r\nu_q) QS
\\
&&{\cal M}_2(s,q^2) = \left( \frac{5\nu}{\nu_q} - 2r -
\frac{\nu}{\nu_q} \frac{s}{\md^2} \right) Q \overline{R}
\\
&&\qquad{}+\frac{r}{2\md} (\nu_p-\md) R^2 - r(r\nu_q + \nu)
\overline{R} S
\\
&&\qquad{}+r(r\nu_q-\nu) R \overline{R} + r^2 \nu_q\nu \overline{R}^2 
\\
&&\qquad{}+\left( \frac{\nu}{\nu_q} \left( 1+ \frac{s}{\md^2}\right) -
\frac{r}{\md} (\nu_p+\md) \right) RS
\\
&&\qquad{}+\left( \frac{3r}2 - \frac{2\nu}{\nu_q} -
\frac{r\nu_p}{2\md} \right) S^2
\\
&&{\cal M}_3(s,q^2) = \left( \nu_p\left( \frac{s}{\md^2} -2 \right)
-3\md \right) Q^2
\\
&&\qquad{}+\md\nu_q \left( \frac{\nu}{\nu_q} \left( \frac{s}{\md^2} -2
\right) + 5r - \frac{4r}{\md} \nu_p \right) QR
\\
&&\qquad{}+\md\nu_q \left( \frac{\nu}{\nu_q} \left( 9- \frac{s}{\md^2}
\right) - 6r + \frac{2s}{\md^2} r \right) QS
\\
&&{\cal M}_4(s,q^2) = \left( \frac{s}{\md} - 4\nu_p - 3\md +
\frac{2s}{\md^2} \nu_p \right) Q\overline{R}
\\
&&\qquad{}+(\nu_p-\md) \left( 3 - \frac{s}{\md^2}\right) R^2
\\
&&\qquad{}+(m_N\nu_q - \md\nu) \left( 5- \frac{2s}{\md^2} \right) R
\overline{R} 
\\
&&\qquad{}+\left(3\md - 2\nu_p - \frac{s}{\md} \right) RS
\\
&&\qquad{}+\nu_q\nu \left( 6m_N - \frac{2s}{\md}r - \frac{\nu_q}{\nu}
(\md + \nu_p) r^2 \right) \overline{R}^2
\\
&&\qquad{}+\left(3\md\nu + \frac{s}{\md^2} (2m_N\nu_q - \md\nu) \right)
\overline{R} S
\\
&&\qquad{}+(\nu_p-3\md) S^2
\\
&&{\cal M}_5(s,q^2) = - \left( 3 - \frac{s}{\md^2} \right) (\md +
\nu_p ) \overline{R}^2
\een
with $r=m_N/\md$ and the sum over spins including the photon, nucleon,
and the $\Delta$.

\vskip .5cm
{\bf Appendix C}
\vskip .25cm

We quote here for completeness the full on-shell Ward-identity for ${\bf
W}^F_{\pi\pi}$.  It contains the $\pi\pi$ forward scattering amplitude
$i{\cal T}_{\pi\pi}$ and the pion-spin averaged $\pi \gamma$
scattering amplitude $i{\cal T}_{\pi \gamma}$ both quoted in
\cite{MASTER}.
\ben
&&\frac1{f_\pi^4} {\bf W}^F_{\pi\pi}(q,k_1,k_2) = 
\\
&&=-(2k_1+q)^2 \sum_{a,b} \epsilon^{a3e} \epsilon^{a3f} {\rm Im}\; 
{\cal T}^{be\to bf}_{\pi\pi}\left( (k_1+q), k_2 \right)
\\
&&{}+ (q\to -q)
\\
&&+ \frac{2}{f_\pi^2} \left(g_{\mu\nu} - (2k_1+q)_\mu k_{1\nu}
\Delta_R(k_1+q) \right) {\rm Im}\; {\cal T}_{\pi\gamma}^{\mu\nu}(q,k_2)
\\
&&+(q\to-q) + (k_1\to-k_1) + (q,k_1\to -q, -k_1)
\\
&&+ \frac{1}{f_\pi} k_1^\alpha (2k_1+q)^\mu \sum_a \epsilon^{a3e}
{\rm Im}\; {\cal A}^{ae}_{\alpha\mu}(k_1,q)
\\
&&+ (q\to -q) + (k_1\to -k_1) + (q,k_1 \to -q,-k_1)
\\
&&+{\rm Im}\; 
\sum_b\frac{3m_\pi^2}{f_\pi^2} \int\! d^4x\, d^4y\; e^{iq\cdot(x-y)}
\\
&&\times\langle \pi^b_{\rm out}(k_2) | T^* {\bf
V}^3_\mu(x) {\bf V}^{3,\mu}(y) \hat{\sigma}(0) | \pi^b_{\rm
in}(k_2) \rangle
\\
&&-\frac{1}{f_\pi^2} \sum_a \epsilon^{a3e} \epsilon^{e3g}
(2k_1+q)^\mu k_1^\alpha \Delta_R(k_1+q) {\rm Im}\;
{\cal B}^{ag}_{\alpha\mu}(k_1,k_2) 
\\
&&- \frac1{f_\pi^2} \left( g^{\mu\nu} - (k_1+q)^\mu (2k_1+q)^\nu
\Delta_R(k_1+q) \right) 
\\
&&\times \sum_a \epsilon^{a3e} \epsilon^{a3f}  {\rm Im}\;
 {\cal B}^{ef}_{\mu\nu}(k_1+q,k_2)   
\\
&&{}-\frac1{f_\pi^2} k_1^\alpha k_1^\beta \Delta_R(k_1) 
\sum_a \epsilon^{a3e} \epsilon^{g3e}
{\rm Im}\; {\cal B}^{ag}_{\alpha\beta}(k_1,k_2)
\\
&&{}+\frac1{f_\pi^2} k_1^\alpha k_1^\beta (2k_1+q)^2 \Delta_R(k_1+q)
\Delta_R(k_1) 
\\
&&\times\sum_a \epsilon^{a3e} \epsilon^{g3e}
{\rm Im}\; {\cal B}^{ag}_{\alpha\beta}(k_1,k_2) 
\\
&&{}+(q\to -q) + (k_1 \to -k_1) + (q,k_1\to -q,-k_1)
\\
&&{}-\frac1{f_\pi^2} (k_1+q)^\alpha (k_1+q)^\beta (2k_1+q)^2
\Delta^2_R(k_1+q) 
\\
&&\times\sum_a \epsilon^{a3e} \epsilon^{a3f} {\rm Im}\; {\cal
B}^{ef}_{\alpha\beta} (k_1,k_2)
+(k_1\to - k_1)
\\
&&- \frac1{f_\pi^2} g^{\mu\nu} k_1^\alpha \sum_a \epsilon^{a3e} 
{\rm Im}\; i {\cal C}^{ea}_{\mu\nu\alpha}(q,k_1+q,k_1)
+(k_1\to -k_1)
\\
&&+{\rm Im}\;
\frac1{f_\pi^2} (2k_1+q)^\mu (k_1+q)^\alpha k_1^\beta \Delta_R(k_1+q)
\\
&&\times\sum_a\epsilon^{a3e} i {\cal C}^{ea}_{\mu\nu\alpha}(q,k_1+q,k_1) 
\\
&&- \frac{1}{f_\pi^2} k_1^\alpha k_1^\beta \int\! d^4x\, d^4y\, d^4z\;
e^{ik_1\cdot(y- x)} e^{iq\cdot z}
\\
&&\times{\rm Im} \; i\langle \pi^b_{\rm out}(k_2) | T^* {\bf
j}_{A\alpha}^a(x) {\bf j}_{A\beta}^a(y) {\bf V}^3_\mu(z) {\bf
V}^{3,\mu}(0) | \pi^b_{\rm in}(k_2) \rangle
\een 
with
\ben
&&{\cal A}^{ae}_{\alpha\mu}(k_1,q) \equiv 
\sum_b \int\! d^4x\, d^4y \; e^{ik_1\cdot x}e^{iq\cdot y} 
\\
&&\times{\rm Im}\;\langle \pi^b_{\rm out}(k_2) |  T^* \left( {\bf 
j}_{A\alpha}^a(x) {\bf V}_\mu^3(y) \right) \pi^e_{\rm in}(0) |
\pi^b_{\rm in}(k_2) \rangle
\\
&&=\frac1{f_\pi}
\delta^{ae} (2k_1+q)_\alpha \int\! d^4x \;e^{ik_1\cdot x} 
\\
&&\times{\rm Im}\; \langle \pi^b_{\rm out}(k_2) | T^* {\bf V}^3_\mu(x)
\hat{\sigma}(0) | \pi^b_{\rm in}(k_2) \rangle
\\
&&{}+{\rm Im}\; \frac1{f_\pi}
\epsilon^{e3g} (2k_1+q)_\mu k_1^\beta \Delta_R(k_1) {\cal
B}^{ag}_{\alpha\beta}(k_1,k_2) 
\\
&&{}+\frac1{f_\pi} \epsilon^{e3g} (\delta^\beta_\mu - (2k_1+q)_\mu
k_1^\beta ){\rm Im}\; {\cal B}^{ag}_{\alpha\beta}(k_1,k_2)
\\
&&{}-\frac1{f_\pi} (k_1+q)^\beta \int\! d^4x\, d^4y\;
e^{iq\cdot x} e^{ik_1\cdot y}
\\
&&\times{\rm Im}\; i\langle \pi_{\rm out}^b(k_2) | T^* {\bf V}^3_\mu(x) {\bf
j}_{A\alpha}^a(y) {\bf j}_{A\beta}^e(0)| \pi^b_{\rm in}(k_2) \rangle
\\
&&{}+ \frac1{f_\pi} \epsilon^{age} \int\! d^4x \; e^{iq\cdot x}
\\
&&\times{\rm Im}\; i\langle \pi^b_{\rm out}(k_2) |  T^* {\bf V}^3_\mu(x) {\bf
V}^g_{\alpha}(0) | \pi^b_{\rm in}(k_2) \rangle
\een
\ben
&&{\cal B}^{ag}_{\alpha\mu}(k_1,k_2) \equiv
\\
&&=i\sum_b \int\! d^4x \; e^{ik_1\cdot x} 
\langle \pi^b_{\rm out}(k_2) | T^*  {\bf
j}_{A\alpha}^a(x) {\bf j}_{A\mu}^g(0) | \pi^b_{\rm in}(k_2)
\rangle 
\\
&&{\cal C}^{ea}_{\mu\nu\alpha}(q,k_1+q,k_1)
\equiv
\\
&&=\sum_b \int\! d^4x\, d^4y \; e^{i(k_1+q)\cdot x} e^{-ik_1\cdot y}
\\
&&\times\langle \pi^b_{\rm out}(k_2) | T^* {\bf V}^3_\mu(0) 
{\bf j}_{A\nu}^e (x) {\bf j}_{A\alpha}^a(y) | \pi^b_{\rm in}(k_2) \rangle
\een

\vskip 1cm
\setlength{\baselineskip}{15pt}

\end{document}